\documentclass[]{llncs}
\usepackage{graphicx}
\usepackage{algorithm}
\usepackage[noend]{algpseudocode}
\usepackage{booktabs}
\usepackage{enumitem} 

\begin{document}
\title{PE-AONT: Partial Encryption combined with an All-or-Nothing Transform}

%
\author{Katarzyna Kapusta \and Gerard Memmi}
\authorrunning{K. Kapusta and G. Memmi}
%
\institute{LTCI, Telecom ParisTech\\
\email{\{katarzyna.kapusta,gerard.memmi\}@telecom-paristech.fr}}
\maketitle              

\begin{abstract}

In this paper, we introduce PE-AONT: a novel algorithm for fast and secure data fragmentation. Initial data are fragmented and only a selected subset of the fragments is encrypted. Further, fragments are transformed using a variation of an all-or-nothing transform that blends encrypted and non-encrypted fragments. By encrypting data only partially, we achieve better performance than relevant techniques including data encryption and straightforward fragmentation. Moreover, when the ratio between the number of encrypted and non-encrypted fragments is wisely chosen, data inside fragments are protected against exposure of the encryption key unless all fragments are gathered by an attacker.

\end{abstract}

\section{Introduction}

Fragmenting and dispersing data over multiple independent storage sites reinforces its confidentiality, integrity, and availability~\cite{depsky}. During the past decades, multiple fragmentation methods were defined and used in various settings ~\cite{future:internet}. These methods include but are not limited to information-theoretic secret sharing schemes~\cite{shamir}, information dispersal algorithms~\cite{rabin}, as well as schemes combining symmetric encryption with data fragmentation~\cite{krawczyk,resch}. 

In modern distributed systems, the most common fragmentation approach is to form final fragments from large chunks of encrypted data. Data are encrypted, fragmented in a straightforward manner and then dispersed over different servers or storage sites. To facilitate key management, encryption keys can also be fragmented using a secret sharing scheme and attached to data fragments. To ensure resilience, data replication or error-correction codes ~\cite{RS} may be applied. This is a reliable way of proceeding, however it does not protect against powerful attacker able to recover encryption keys. Indeed, such attackers may obtain a portion of information by just decrypting fragments in their possession.

Protection against key exposure requires additional post-processing that will create dependencies between ciphertext blocks contained inside the fragments~\cite{bastion} or that will shred ciphertext blocks over the fragments~\cite{trustcom}. Encrypted data are then protected against key exposure unless all the fragments are gathered. Such additional protection comes at a performance price, as it increases the required number of operations.

On another hand, partial encryption is usually used in lightweight cipher. In this paper, we introduce PE-AONT: a fragmentation algorithm that is not only faster than the most common fragmentation technique, but also (when wisely applied) protects encrypted data inside fragments against a situaion of key exposure. To achieve this, PE-AONT fragments initial data into $k$ fragments, but encrypts only $e$ of them. In a next step, it applies an all-or-nothing transform over the totality of the fragments that exclusive-ors ciphertext and plaintext blocks. PE-AONT is then combining AONT technique with partial encryption achieving speed and a good level of security. A secure dispersal of fragments ensures that data from a single fragment cannot be recovered unless all fragments are being gathered. Limiting data encryption allows to speed up the performance, while the all-or-nothing transform combined with fragmentation ensures good level of data protection. By adjusting the values of $k$ and $e$, a user may balance between the desired performance and data protection level.

\noindent\textbf{Outline} In Section~\ref{sec:relevant} we present state-of-the-art techniques for fast data fragmentation. In Section~\ref{sec:description} we describe in details the proposed PE-AONT algorithm. In Section~\ref{sec:comparison} we compare PE-AONT with relevant works, analyzing its complexity and level of data protection. Section~\ref{sec:performance} contains the performance benchmark. We conclude with an insight into future works.

\section{Relevant works}
\label{sec:relevant}

This section presents relevant fragmentation techniques. During descriptions we will use the following notation. Initial data $d$ are transformed into a set of $n$ fragments, $k$ of which are needed for data reconstruction. In Bastion scheme, a sequence of $l$ ciphertext blocks $x=X_1,\ldots,X_l$ is transformed into a sequence of $l$ $x'=X'_1,\ldots,X'_l$ transformed blocks.

\subsection{Encryption and Straightforward Fragmentation}

 A computationally secure scheme for fragmentation of larger data was first introduced by Krawczyk in his seminal work \textit{Secret Sharing Made Short} (SSMS) ~\cite{krawczyk}. SSMS can be considered as a more general methodology combining three elements: data encryption, data fragmentation, and secret sharing of the encryption key. In the original proposal, data $d$ is encrypted using a symmetric encryption algorithm, and then fragmented using an information dispersal algorithm into $n$ fragments, $k$ of which are needed for the data recovery. Encryption key is split using a perfect secret sharing scheme (typically Shamir's secret sharing~\cite{shamir}) and dispersed within data fragments. In contrast to perfect secret sharing schemes (like the one-time pad or Shamir's scheme), the storage overhead of SSMS does not depend on data size, but is equal to the size of the key per data fragment (for larger data it is practically negligible).
 
 Performance of SSMS depends on implementation details of encryption and dispersal techniques. In modern implementations, systematic error-correction codes are used as the dispersal technique~\cite{depsky,RS}. This improves the performance, but allows a partial decryption of compromised fragments in a situation of key exposure, as the set of fragments contains fragments that are formed from large encrypted data chunks.
 
 Alike SSMS, the AONT-RS method~\cite{revisiting-aont-rs,resch} combines symmetric encryption with data dispersal. The difference between those two methodologies lies in the key management. In AONT-RS, the key is  exclusive-ored with the data hash.

\subsection{Bastion scheme}

Recently introduced, Bastion~\cite{bastion} transforms a ciphertext in order to make its decryption computationally infeasible unless all but two ciphertext blocks are being gathered. In more details, the sequence of $l$ input ciphertext blocks $x=X_1,\ldots,X_l$ is multiplied by a square matrix $A$, such that: (i) all diagonal elements are set to 0, and (ii) the remaining off-diagonal elements are set to 1 (such matrix is invertible and $A=A^{-1}$, so the inverse transform: $x=A^{-1}\cdot x'=A \cdot x'$). The multiplication $x'=A\cdot x$ ensures that each output block $X'_i$ will depend on all output blocks $X'_j$ except from $X'_i$,$i=j$. Bastion achieves much better performance than other all-or-nothing transforms as it requires only $2l$ exclusive-or operations in addition to data encryption.

\subsection{Secure Dispersal and Fragmentation}

Secure fragmentation and dispersal (SFD)~\cite{trustcom} divides a ciphertext encrypted using a block cipher with a mode of operation that creates chaining between consecutive ciphertext blocks (like Cipher Block Chaining) into fragments resisting to key leakage. The complete  scheme is composed of three steps. The first step separates consecutive blocks of the ciphertext. The second step separates bits of blocks over final fragments. At last, fragments are dispersed and stored over independent storage locations e.g. clouds. In consequence, an attacker present at a single storage location is unabl e to decrypt a single block of the ciphertext even if she possesses the right encryption key. SFD can be easily integrated within SSMS or AONT-RS methodologies in order to reinforce their protection level by providing additional protection against key exposure. 

\section{Description of the algorithm}
\label{sec:description}

PE-AONT is composed of three steps. In a first step, initial data $d$ composed of $m$ plaintext blocks are fragmented into $k$ fragments, where only $e$ of the $k$ fragments are encrypted. In a second step, the $k$ fragments are transformed using an all-or-nothing transform that blends ciphertext contained inside the $e$ encrypted fragments with the $k-e$ plaintext fragments. Both steps are described in details in the following subsections.

\subsection{Step1: Data fragmentation and partial encryption}

\begin{figure}
\includegraphics[width=0.99\linewidth]{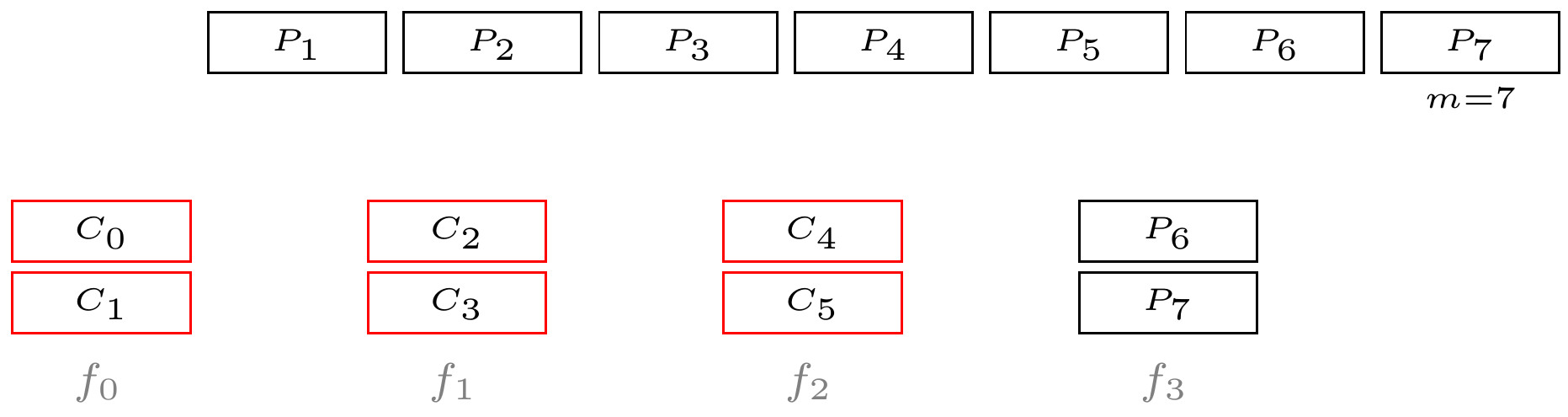}
\caption{\textit{Step 1, example for $k=4$, $e=3$, and $m=7$. First step of the algorithm fragments data composed of $m$ blocks into $k$ fragments of size $\frac{l}{k}=2$, where $l=m+1$ blocks each. The block $C_0$ is added to $d$; it corresponds to the ciphertext block containing the Initial Vector used by the cipher.}}
\label{fig:example-step1}
\end{figure}

Pseudo-code of the first step of the algorithm - \textsc{Fragment\-And\-Encrypt}- is presented in Figure~\ref{fig:step1}. An example illustrating the process is shown in Figure~\ref{fig:example-step1}.
During this step, initial data $d$, composed of $m$ plaintext blocks, is fragmented into $k$ fragments. Only $e<k$ of the fragments are being encrypted using a block cipher algorithm. During the encryption process one more block is inserted containing the initial vector. Therefore, at the end of the processing, each fragment contains $\#f=\frac{l}{k}=\frac{m+1}{k}$ (plaintext or ciphertext) blocks.

\begin{figure}[h!]
\begin{algorithmic}[1]
\Function{FragmentAndEncrypt}{$d,e,k$}
\State Fragment data $d$ into $k$ initial fragments $f_0,\ldots,f_{k-1}$
\State Encrypt $e$ of the $k$ fragments:
    \For {each fragment $f_i$, $i=0,\ldots,e-1$}  
	\State Encrypt fragment $f_i$
    \EndFor
\EndFunction
\end{algorithmic}

\caption{\textit{Pseudo-code of Step 1 fragmenting data into $k$ fragments.}}
\label{fig:step1}
\end{figure}

\subsection{Step 2: All-or-nothing transform of the fragments}

\begin{figure}
\includegraphics[width=0.99\linewidth]{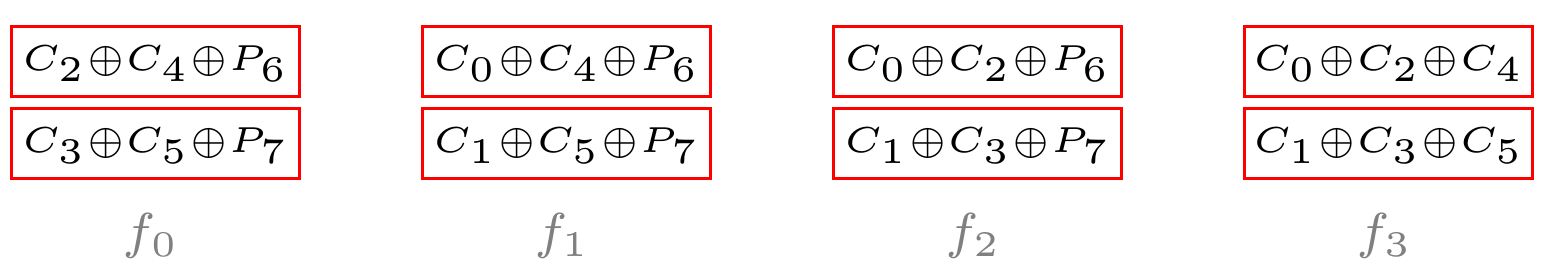}
\caption{\textit{Step2, example for $k=4$ and $e=3$. Fragments are transformed using an all-or-nothing transform that exclusive-ors ciphertext and plaintext blocks. The transform is applied 'row' by 'row' over $k$ blocks (each block comes from a different fragment) protecting the $k-e$ fragments which were not encrypted.}}
\label{fig:example-step2}
\end{figure}

The pseudo-code of the second step - \textsc{AONT\-Transform} - is presented in Figure~\ref{fig:step2}. An example illustrating the process is presented in Figure~\ref{fig:example-step2}. Blocks inside of the $k$ fragments are processed simultaneously, from the block $0$ to $\#f-1$, by sets of $k$ elements (in a \textit{row by row} fashion, each block comes from a different fragment). The all-or-nothing transform presented in \cite{bastion,stinson} is applied over the set of $k$ blocks, $e$ of which are ciphertext blocks and the $k-e$ remaining ones are plaintext blocks.

\begin{figure}[h!]
\begin{algorithmic}[1]
\Function{AONTransform}{$f$}
    \For {$i=0,\ldots,\#f-1$}  
	\State $sum = \bigoplus_{0}^{j=k-1}C_{\#f\times j+i}$
	\For {$j=0,\ldots,k-1$} 
    \State $C_{\#f\times j+i}=C_{\#f\times j+i}\oplus sum$
	\EndFor
    \EndFor
\EndFunction
\end{algorithmic}
\caption{\textit{Pseudo-code of the Step 2. An all-or-nothing transform is applied over the $k$ fragments exclusive-oring their plaintext and ciphertext blocks. $\#f$ denotes the number of blocks inside a fragment. Blocks are processed by 'rows' of $k$ blocks (each block from a different fragment). For each 'row' of $k$ blocks a $sum$ value containing the exclusive-or of the $k$ blocks is first computed. $sum$ is then exclusive-ored with each value in the \textit{row}. }}
\label{fig:step2}
\end{figure}

\subsection{Step 3: Dispersing fragments}

In a final step, fragments are dispersed over independent storage sites. The dispersal technique depends on values of $e$ and $k$. 
\begin{itemize}[leftmargin=*]
\item For $e=k-1$: no more than $k-2$ fragments can be stored at a single storage site. This comes from the property of the linear all-or-nothing transform protecting transformed blocks unless more than $k-2$ blocks are exposed~\cite{bastion}.
\item For $e<k-1$: all fragments have to be dispersed over independent storage sites. As the number of plaintext fragments increases, it is possible that same combinations of ciphertext blocks will be used to protect different plaintext blocks. Indeed, data protection level is lower and fragments are only secure when the assumed attacker is not able to compromise more than one storage site. It is the price that has to be paid for speeding up the algorithm processing.
\end{itemize}

\begin{table}[h]
  \caption{\textit{Comparison in terms of number of block cipher operations (block op.), number of exclusive-ors, and the ability to provide a key exposure protection (K.E.P). We assume a plaintext of $m=l-1$ blocks. (*) PE-AONT provides key exposure protection when $e\ge3$. When $e<k-1$ fragments are protected against key exposure, but data is protected only against an attacker at a single storage site.}}
  \label{tab:comparison}
  \begin{tabular}{llll}
    \toprule
    Algorithm & Block op. & Exclusive-ors & K.E.P.\\
    \midrule
    Encryption & l-1 b.c. & l-1  & No \\
    Bastion & l-1 b.c. & 3l-1 [AON Transform: 2l]  & Yes \\
    SFD & l-1 b.c. & l-1  & Yes \\
    \textbf{PE-AONT} & $e\frac{l}{k}$ b.c. & $e\frac{l}{k} + 2l$ &  \textbf{Yes*}  \\
  \bottomrule
\end{tabular}
\end{table}

\section{Theoretical Comparison}
\label{sec:comparison}

We compared PE-AONT with relevant works in terms of amount of computations and ability to protect against key exposure. Results are shown in Table~\ref{tab:comparison}. As a baseline, we use encryption that requires $l-1$ block cipher operations and $l-1$ exclusive-or operations (when applied on a plaintext composed of $m=l-1$ blocks). Bastion scheme applies only a linear transform over the encrypted data without increasing the number of block cipher operations. Bastion's transform uses $2l$ exclusive-or operations. Counting with the encryption step, Bastion scheme requires $3l-1$ exclusive-ors. SFD does not require additional operations in addition to data encryption, as it just disperses data chunks over fragments (depending on the programming technique this also leads to a slight performance overhead of up to 11\%). PE-AONT requires only $e\frac{l}{k}$ block cipher operations, as it does not encrypt the totality of the data contained inside the fragments. It performs $2l + e\frac{l}{k}$ exclusive-or operations: $ e\frac{l}{k}$ during the encryption of part of the data and $2l$ during the AONT. 

Encryption and straightforward fragmentation do not protect against key exposure: an attacker possessing the key is able to decrypt the transformed data. Bastion protects transformed ciphertext against key exposure unless all but two blocks are being exposed. The level of data protection in PE-AONT depends on the values of $e$ and $k$. To achieve the protection against key exposure, each plaintext block should be exclusive-ored with at least two different ciphertext blocks. The minimum value of $k$ is then $4$ (as the number of input blocks in the AONT transform has to be even, see the explanation in:~\cite{bastion}) and the minimum value of $e$ is $3$. Intuitively, the larger the value of $e$, the higher will be the provided level of data protection. When $e=k-1$, data cannot be decrypted unless all $k$ fragments are gathered. Fragments are thus protected against key exposure. When $3\ge e\le k-2$ the protection is lower and ensured only if the attacker is at a single storage site. Indeed, when the attacker compromised more than one site, she can obtain relationships between fragments by exclusive-oring the fragments in her possession (a similar problem occurs when a one-time pas is reused). Therefore, the most interesting configuration of PE-AONT is the one when $e=k-1$, as it is not only faster than encryption and straightforward fragmentation, but also protects data against key exposure.

\section{Performance comparison with relevant works}
\label{sec:performance}

\textbf{Implementation details} Relevant algorithms were implemented using the same programing style in JAVA with JDK 1.8 on DELL Latitude E6540, X64-based PC running on Intel\textsuperscript{\textregistered} Core\textsuperscript{TM} i7-4800MQ CPU @ 2.70 GHz with 8 GB RAM, under Windows 7. Standard $javax.crypto$ library was used.  A random data sample was used for each measurement and each presented result is an average of 30 measurements. AES-NI with 128 bits key was used for encryption. 

\begin{figure}
\includegraphics[width=0.99\linewidth]{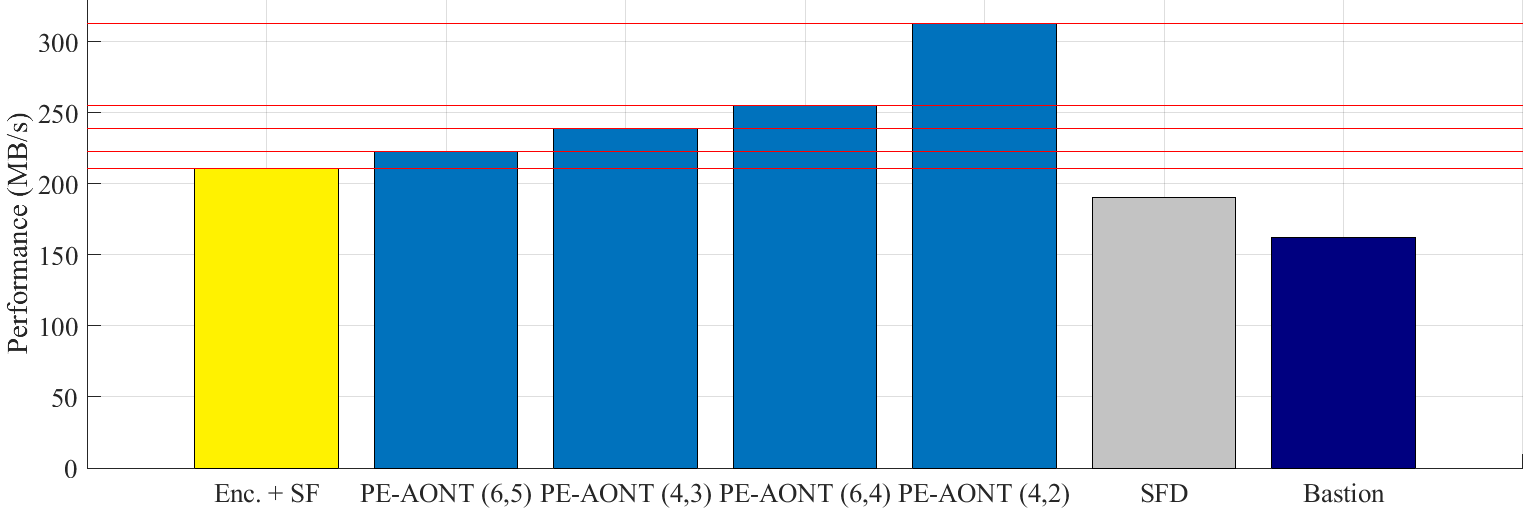}
\caption{\textit{Performance benchmark. PE-AONT was measured in different configurations of $(k,e)$. In all of them it is faster than encryption and straightforward fragmentation. It is also faster than two fragmentation schemes protecting against data exposure: secure fragmentation and dispersal (SFD) and the Bastion scheme.}}
\label{fig:performance}
\end{figure}

Performance comparison between relevant algorithms is presented in Figure~\ref{fig:performance}. The performance of PE-AONT is shown in 4 configurations: two configurations were $e=k-1$ (and where the protection level is high) and for two configurations when $e\leq k-2$ (data protection is traded for better performance). In all configurations, PE-AONT outperforms encryption and straightforward fragmentation. Moreover, it is much faster than two relevant fragmentation techniques protecting fragmented data against key exposure: SFD and Bastion's scheme.

\section{Future works}
\label{sec:future}

Our software integrates encryption and  PE-AONT in a coarse-grained fashion. In the future, we would like to integrate the algorithm within the encryption process to allow an even better fragmentation performance. We recently improved the Bastion AONT algorithm by decreasing the number of Xor operations needed to protect data against key exposure in ~\cite{ccs} or in ~\cite{css}. it would be interesting to analyze and compare in which situation one method is preferable over another one. PE-AONT is very likely best fitted as a lightweight cipher since it is faster than the other AONT algorithms. Moreover, we would like to integrate PE-AONT inside a complete distributed system protecting data using a combination of fragmentation, encryption, and dispersal~\cite{kapusta:shanghai}. 

\section{Conclusion}

PE-AONT: a fast fragmentation method combining partial encryption with an all-or-nothing transform was introduced in order to speed up the fragmentation performance. The desired performance and protection levels can be adjusted by the user who can set parameter values to increase performance and decrease protection or the other way around. An experimental evaluation shows that the algorithm achieves better performance than relevant techniques including the most common way of data fragmentation. As such, it can be chosen as a lightweight cipher. Moreover, by carefully choosing parameters values, it protects fragmented data against a situation of key exposure as long as an attacker does not access to the entire set of fragments. We now believe that PE-AONT could be considered as a software component and successfully be integrated within modern distributed systems where rapid protection is required.

\bibliographystyle{splncs04}


\end{document}